\documentclass[a4paper,oneside,reqno,11pt]{amsart}
\usepackage{amssymb}
\usepackage[foot]{amsaddr}
\usepackage{pdflscape}
\usepackage{url}
\usepackage{cite}
\usepackage{tikz}
\usepackage{booktabs}
\usepackage{caption}
\usepackage{float}
\usepackage[bottom=3cm,top=3cm]{geometry}
\usepackage{hyperref}
\definecolor{darkblue}{RGB}{0,0,127} 
\definecolor{darkgreen}{RGB}{0,150,0}
\hypersetup{breaklinks, colorlinks, linkcolor=darkblue, citecolor=darkgreen, filecolor=red, urlcolor=blue}

\theoremstyle{definition}
\newtheorem{fact}{Fact}
\newcommand{\ea}[1]{\begin{align}#1\end{align}}
\newcommand{\nn}{\nonumber \\}
\newcommand{\vpu}[1]{^{\vphantom{#1}}}
\newcommand{\bc}{\begin{center}}
\newcommand{\ec}{\end{center}}
\newcommand{\bmt}{\begin{pmatrix}}
\newcommand{\emt}{\end{pmatrix}}

\newcommand{\mbf}[1]{\mathbf{#1}}
\newcommand{\mcl}[1]{\mathcal{#1}}
\newcommand{\fd}[1]{\mathbb{#1}}
\newcommand{\la}{\langle}
\newcommand{\ra}{\rangle}

\DeclareMathOperator{\Tr}{Tr}

\DeclareMathOperator{\Rank}{Rank}
\DeclareMathOperator{\gal}{Gal}

\DeclareMathOperator{\EC}{EC}

\DeclareMathOperator{\PEC}{PEC}
\DeclareMathOperator{\CEC}{CEC}

\DeclareMathOperator{\ESL}{ESL}
\DeclareMathOperator{\GL}{GL}

\DeclareMathOperator{\stbz}{S_0}
\DeclareMathOperator{\stb}{S}

\title{SICs and Algebraic Number Theory}

\author{$^*$Marcus Appleby}
\email{marcus.appleby@sydney.edu.au}
\author{$^{*,\S}$Steven Flammia}
\email{steven.flammia@sydney.edu.au}
\author{$^\dagger$Gary McConnell}
\email{g.mcconnell@imperial.ac.uk}
\author{$^\ddagger$Jon Yard}
\email{jyard@uwaterloo.ca}

\address{$^*$Centre for Engineered Quantum Systems, School of Physics, University of Sydney, Sydney, Australia}
\address{$^\S$Center for Theoretical Physics, Massachusetts Institute of Technology, Cambridge, USA}
\address{$^\dagger$Controlled Dynamics Theory Group, Imperial College, London, UK}
\address{$^\ddagger$Institute for Quantum Computing, Dept. of Combinatorics and Optimization, University of Waterloo and Perimeter Institute for Theoretical Physics, Waterloo, Canada} 

\begin{document}

\begin{abstract}

We give an overview of some remarkable connections between symmetric informationally complete measurements (SIC-POVMs, or SICs) and algebraic number theory, in particular, a connection with Hilbert's $12^{\rm{th}}$ problem. 
The paper is meant to be intelligible to a physicist who has no prior knowledge of either Galois theory or algebraic number theory. 
\end{abstract}

\maketitle

\allowdisplaybreaks

\section{Introduction}
Symmetric informationally complete measurements (SIC-POVMs, or SICs as we call them here) are a special type of quantum measurement that was originally introduced in refs.~\cite{Zauner:1999,Renes:2004}. 
They possess, as the name would suggest, a particular symmetry that is natural and elegant. 
They have numerous applications in quantum information~\cite{Fuchs:2003,Rehacek:2004,Englert:2004,Scott:2006,Durt:2008,Zhu:2011,Zhu:2016a}, quantum foundations~\cite{Fuchs:2013,Fuchs:2016a,Appleby:2016b} and classical signal processing~\cite{Howard:2006,Malikiosis:2016}. 
They have also been implemented experimentally~\cite{Du:2006,Durt:2008,Medendorp:2011,Pimenta:2013}. 
However, their existence has only been established in a sizable, but nonetheless finite, number of cases: 
exact solutions\cite{Scott:2010a,Appleby:2017} have been calculated in dimensions $2$--$21$, $24$, $28$, $30$, $35$, $39$, $48$ and numerical solutions\cite{Scott:2010a,Scott:2016} have been calculated in every dimension up to 139, and for a handful of other dimensions up to 323. 
This encourages the speculation that they exist in every finite dimension; however, in spite of a great deal of theoretical effort, a proof continues to elude us.

A SIC is a geometrical structure. 
However, in this paper we are going to put the geometry on one side and focus instead on the numbers: i.e.\ the matrix elements of the fiducial projectors, and the overlap phases. 
To someone coming to the problem for the first time the numbers may seem rather off-putting. 
Since a SIC is defined by its simple symmetry properties, one might expect that there should be correspondingly simple expressions for the numbers. 
This is true of mutually unbiased bases (MUBs), for example~\cite{Ivanovic:1981,Wootters:1987}. 
However, it seems on the face of it that it is typically not the case for SICs. 
It is true that the numbers appearing in the $d=4$ SICs are comparatively simple (a fact which is exploited in the paper by Bengtsson~\cite{Bengtsson:2016} in this volume). 
But the complexity of the solutions increases rapidly with dimension, to such an extent that the exact fiducial projector on orbit $48a$ calculated in ref.~\cite{Appleby:2017} occupies $\sim 50,000$ A4 pages of print-out in font 9, and with narrow margins -- this is $\sim 20$ pages per individual matrix element! 
We know of at least one colleague who holds that a problem which generates numbers like this cannot be interesting.
 
It is to be observed, however, that beauty is in the eye of the beholder. 
In particular, it can be very sensitive to the perspective, and background knowledge of the beholder. 
For example, if a number is presented as the infinite decimal expansion $1.4142135623730950488016887242096980785696718753769\dots$ it may elicit one aesthetic response, but if presented in the form $\sqrt{2}$ it may elicit quite another. 
The question arises, whether we are looking at the numbers in the SIC problem from the wrong angle. 
Perhaps if we could penetrate their secret the numbers, so far from appearing off-putting, would come to seem the extreme opposite. 
In the following we hope to convince the reader that that is indeed the case.

Specifically, we will describe a recently discovered~\cite{Appleby:2013,Appleby:2016} connection between SICs and some major open questions in algebraic number theory, including Hilbert's $12^{\rm{th}}$ problem. 
The connection begins to explain why the problem of proving SIC existence seems so difficult. 
It also suggests some new approaches to the problem. 
Going in the reverse direction, it may be that SICs have significant applications to number theory. 
They are thus one of a handful of areas spanning both algebraic number theory and quantum information (other examples being quantum compilers~\cite{Kliuchnikov:2015a,Ross:2014, Yard:2016} and quantum algorithms~\cite{Hallgren:2005, Hallgren:2007, Eisentrager:2014}).

%

Rather than proving new results our aim in the following is to make the known results more widely accessible. 
Ref.~\cite{Appleby:2013} made few concessions to the non-expert reader, while ref.~\cite{Appleby:2016} was aimed squarely at a pure mathematical audience. We have received a number of requests to write an account which would be digestible by a physicist who starts out knowing little or nothing about algebraic number theory. 
This paper is our response to those requests. 
It is aimed at the same target audience as the paper by Bengtsson~\cite{Bengtsson:2016}, also in this volume, and it could usefully be read in conjunction with that paper.

We begin in Section~\ref{sec:SICs} with a brief summary of the basic properties of a SIC. The next six sections are arranged in pairs. 
The first section in each pair describes an aspect of number theory in a way that is hopefully intelligible to the non-expert reader, while the second explains its relation to the SIC problem. 
The first such pair comprises Section~\ref{sec:Galois}, which describes the basic facts of Galois theory, and Section~\ref{sec:GaloisSIC}, which describes the Galois group of a SIC. 
The second pair comprises Section~\ref{sec:Hilbert}, which describes the mathematical background to Hilbert's twelfth problem, and Section~\ref{sec:SICsRayClass}, which describes how SICs generate certain of the ray class fields featuring in the problem. 
The third pair comprises Section~\ref{sec:AlgInt}, which describes the ring of algebraic integers and unit group associated to an algebraic number field, and Section~\ref{sec:SICsUnit}, which describes the subgroup of the unit group generated by the SIC overlap phases. 
Finally, in Section~\ref{sec:conc} we discuss the implications of these results.

\section{SICs: Basic Properties}
\label{sec:SICs}
The purpose of this section is to fix notations and summarize some essential facts. 
The treatment is correspondingly terse. 
The discussion will become more leisurely in subsequent sections. 
For more details regarding the material in this section see refs.~\cite{Appleby:2005b,Scott:2010a}. 

A SIC in dimension $d$ is a family of $d^2$ rank-$1$ projectors $\Pi_1, \dots, \Pi_{d^2}$ satisfying
 \ea{
 \Tr(\Pi_r\Pi_s) &= \frac{d\delta_{r,s}+1}{d+1}
 }
Every known SIC has a group covariance property. 
That is, there exists a group of unitary operators $U_1, \dots, U_n$ which permute the SIC projectors according to
\ea{
U_j \Pi_r U^{\dagger}_j &= \Pi_{\sigma_j(r)}
} 
(where $\sigma_j$ is a permutation of the integers $1, \dots, d^2$), in such a way that the action is transitive. 
The transitivity of the action means that in order to specify the SIC it is enough to specify a single projector, which we call the fiducial projector.

Zhu~\cite{Zhu:2010} has shown that in prime dimension $d$, if a SIC has a group covariance property at all, then it is necessarily covariant with respect to the $d$-dimensional Weyl-Heisenberg group. 
In practice it seems that, with the single exception of the Hoggar lines~\cite{Hoggar:1998,Zhu:2015a,Stacey:2016} in dimension 8, SICs are always covariant with respect to this group even when $d$ is not a prime. 
From now on we will exclude the Hoggar lines from consideration, and will use the term ``SIC'' specifically to mean a SIC which is covariant with respect to the $d$-dimensional Weyl-Heisenberg group. 

We now describe the Weyl-Heisenberg group. 
Let $|0\ra, \dots, |d-1\ra$ be the standard basis in dimension $d$ and let $X$, $Z$ be the unitaries which act according to
\ea{
X|r\ra &= |r+1\ra & Z|r\ra &= \omega^r |r\ra
}
where the addition in the expression $|r+1\ra$ is $\mod d$ and $\omega = \mathrm{e}^{\frac{2\pi i }{d}}$. 
We then define the Weyl-Heisenberg displacement operators by 
\ea{
D_{\mbf{p}} &= \tau^{p_1p_2} X^{p_1} Z^{p_2}
}
where $\mbf{p}$ is the ``vector'' $\left(\begin{smallmatrix} p_1 \\ p_2 \end{smallmatrix}\right)$ and $\tau = -\mathrm{e}^{\frac{\pi i}{d}}$. 
The Weyl-Heisenberg group is then the group generated by the operators $D_{\mbf{p}}$. 
The necessary and sufficient condition for a rank-$1$ projector $\Pi$ to be the fiducial projector for a Weyl-Heisenberg covariant SIC is that
\ea{
\Tr(\Pi D_{\mbf{p}}) &= \frac{\mathrm{e}^{i\theta_{\mbf{p}}}}{\sqrt{d+1}}
\label{eq:olpDef}
}
for all $\mbf{p}\neq \boldsymbol{0} \mod d$ and some set of phases $\mathrm{e}^{i\theta_{\mbf{p}}}$. 
In the sequel the numbers $\Tr(\Pi D_{\mbf{p}})$ will be called the overlaps, and the numbers $\mathrm{e}^{i\theta_{\mbf{p}}}$ the overlap phases. 
The displacement operators satisfy
\ea{
\Tr(D\vpu{\dagger}_{\mbf{p}}D^{\dagger}_{\mbf{q}}) &= d \delta_{\mbf{p},\mbf{q}}.
}
They consequently form an orthogonal basis for operator space relative to the Hilbert-Schmidt inner product. 
It follows that the overlaps completely determine the fiducial projector $\Pi$. 

The extended Clifford group also plays an important role in the theory. 
Let $d'=d$ (respectively $d'=2d$) if $d$ is odd (respectively even) and define the extended symplectic group $\ESL(2,\fd{Z}/d'\fd{Z})$ to consist of all $2\times 2$ matrices with entries in $\fd{Z}/d'\fd{Z}$ and determinant $\pm 1$. 
For each $F \in \ESL(2,\fd{Z}/d'\fd{Z})$ there exists an operator $U_F$, unique up to an overall phase, such that 
\ea{
U\vpu{\dagger}_F D\vpu{\dagger}_{\mbf{p}} U^{\dagger}_F = D_{F\mbf{p}}.
}
The operator $U_F$ is unitary (respectively anti-unitary) if $\det F = +1$ (respectively $\det F = -1$). 
The map $F \to U_F$ is a projective representation of $\ESL(2,\fd{Z}/d'\fd{Z})$. 
So
\ea{
U_{F_1} U_{F_2}&\dot{=} U_{F_1 F_2} & \forall F_1, F_2 &\in \ESL(2,\fd{Z}/d'\fd{Z})
\\
U\vpu{\dagger}_{F^{-1}}& \dot{=} U^{\dagger}_{F} & \forall F & \in \ESL(2,\fd{Z}/d'\fd{Z})
}
where $\dot{=}$ means ``equal up to a phase''. 
If $d$ is an odd prime the operators $U_F$ can be chosen in such a way that the map becomes an ordinary, non-projective representation~\cite{Appleby:2009}. 
The extended Clifford group $\EC(d)$ is then defined to be the set of all operators of the form $\mathrm{e}^{i\theta} D_{\mbf{p}} U_F$, where $\mathrm{e}^{i\theta}$ is an arbitrary phase. 
Its significance for us is that it preserves ``SICness'': if $\Pi$ is a SIC fiducial, and $U\in \EC(d)$, then $U\Pi U^{\dagger}$ is another Weyl-Heisenberg SIC fiducial. 

If $U, U' \in \EC(d)$ are equal up to a phase, then $U\Pi U^{\dagger} = U' \Pi U^{\prime \dagger}$ for all $\Pi$. So the phases in the definition of $\EC(d)$ are for our purposes irrelevant. 
It is therefore convenient to get rid of them, by defining the projective extended Clifford group
\ea{
\PEC(d) = \EC(d)/\CEC(d)
}
where $\CEC(d)$ is the centre of $\EC(d)$, consisting of all unitaries of the form $\mathrm{e}^{i\theta} I$. 
Note that, unlike $\EC(d)$, the group $\PEC(d)$ is finite. 
Providing the practice can lead to no confusion, we will make no notational or terminological distinction between an operator in $\EC(d)$, and the corresponding coset in $\PEC(d)$. 
For example, we will say that a unitary $U$ belongs to $\PEC(d)$, when what we actually mean is that the coset $U\CEC(d)$ belongs to $\PEC(d)$. 

 The fact that the group $\PEC(d)$ preserves SICness means that one can classify SICs by the $\PEC(d)$ orbit to which they belong. 
With the exception of dimension $3$ it seems that there are only finitely many $\PEC(d)$ orbits in any given dimension. 
Following Scott and Grassl~\cite{Scott:2010a} we label the orbits by the dimension followed by a letter. 
Thus $8a$, $8b$ are the two known orbits of Weyl-Heisenberg SICs in dimension 8 ; $48a$, \dots, $48j$ are the ten known orbits in dimension 48.

We define the stability group of the fiducial projector $\Pi$ to be the set of all $U \in \PEC(d)$ such that $U\Pi U^{\dagger} = \Pi$. 
In every known case it contains a unitary of the form $D_{\mbf{p}} U_F$, where $\Tr(F) = -1 \mod d$, and $F\neq I$ (note that if $d\neq 3$, the second condition is a consequence of the first). Such unitaries are necessarily order $3$, and will be referred to as canonical order $3$. 
Every known $\PEC(d)$ orbit contains fiducials for which the stability group consists entirely of elements of $\ESL(2,\fd{Z}/d'\fd{Z})$. 
We refer to such fiducials as ``centred''. 
For a centred fiducial $\Pi$ we define $\stbz(\Pi)$ to be the group consisting of all $F \in \ESL(2,\fd{Z}/d'\fd{Z})$ such that $U_F$ is in the stability group. 
If $d\neq 3 \mod 9$ every known orbit contains a centred fiducial for which the canonical order $3$ unitary is $U_{F_z}$, where
\ea{
F_z &= \bmt 0 & d-1 \\ d+1 & d-1 \emt
}
is the Zauner matrix.
If $d=3 \mod 9$ the set of orbits split into two disjoint subset. 
Orbits in one subset contain centred fiducials for which the canonical order $3$ unitary is $U_{F_z}$; orbits in the other contain centred fiducials for which the canonical order $3$ unitary is $U_{F_a}$, where
\ea{
F_a &= \bmt 1 & d+3 \\ \frac{4d}{3} & d-2\emt.
}
We refer to these two kinds of orbit as type-$z$ and type-$a$ respectively. Note that if $d$ is even $F_z$, $F_a$ are order 6; however $U_{F_z}$, $U_{F_a}$ are always order 3 (as elements of $\PEC(d)$). 

The known SICs in dimensions $2$ and $3$, along with the Hoggar lines in dimension $8$ have some special properties. 
Following Stacey~\cite{Stacey:2016} we refer to them as sporadic SICs. 
Every other SIC we refer to as generic. 
Special properties of the sporadic SICs include
\begin{enumerate}
\item Some of the sporadic SICs are supersymmetric  in the sense\footnote{This is not the sense of the word as it is used in quantum field theory, to refer to a symmetry relating bosons and fermions, or in what is usually meant by supersymmetric quantum mechanics.} of ref.~\cite{Zhu:2015}.
This is not the case for any generic SIC.
\item The stability groups for the sporadic SICs are non-Abelian, whereas for every generic SIC the stability group is not only Abelian, but even cyclic.
\item In dimension $d$ every known generic SIC is covariant with respect to the $d$-dimensional Weyl-Heisenberg group; this is not the case for every sporadic SIC (namely, the Hoggar lines).
\item The standard basis matrix elements of every known generic SIC projector are algebraic numbers; this is not the case for every sporadic SIC projector.
\item The Galois group of the field associated to every known generic SIC is non-Abelian; this is not the case for every sporadic SIC.
\end{enumerate}

\noindent For the sake of simplicity we will therefore confine ourselves to generic SICs in the following.

\section{Mathematical Background: Galois Theory}
\label{sec:Galois}
The first step in our numerical journey is to note that in every known example of a generic SIC the numbers are expressible in radicals (i.e.\ they can be built up from the integers iteratively, using the four arithmetical operations together with the operation of taking roots). 
It follows that the associated Galois group must be solvable~\cite{Roman:2005}. 
In the next section we give a detailed description of the Galois group, and its action on the overlaps. 
However, before doing so, we review some of the basics facts about Galois theory. 
The discussion will hopefully be intelligible to someone with no prior knowledge of the subject. 
Unfortunately, space does not permit a very extensive treatment. 
For more details the reader may consult one of the standard texts---for example, Roman~\cite{Roman:2005}. 

Perhaps the simplest instance of a Galois group is the Galois group of $\fd{C}$ over $\fd{R}$. 
The complex numbers can be obtained from the reals by appending the single number $i$; a fact which is succinctly conveyed by writing $\fd{C} = \fd{R}(i)$. 
Every number $z \in \fd{C}$ can be written $z= c_0 + c_1 i$, for unique numbers $c_0$, $c_1\in \fd{R}$. 
The Galois group of $\fd{C}$ over $\fd{R}$, denoted $\gal(\fd{C}/\fd{R})$, is the set of all bijections $f\colon \fd{C} \to \fd{C}$ such that
\ea{
f(u + v) &= f(u) + f(v) & \forall u,v & \in \fd{C}
\\
f(uv) &= f(u) f(v) & \forall u,v & \in \fd{C}
\\
f(u) &= u &\forall u &\in \fd{R}
}
Such bijections are called field, or Galois automorphisms.
It follows from the above that 
\ea{
f(c_0+c_1i) = c_0 + c_1 f(i)
}
for all $c_j \in \fd{R}$. 
Consequently $f$ is completely fixed by its action on the generator, $i$. 
To determine the possible values of $f(i)$ observe that we must have $\big(f(i)\big)^2 = -1$ , implying $f(i) = \pm i$. 
So $\gal(\fd{C}/\fd{R})$ contains exactly two automorphisms: The identity and complex conjugation.

In the following we will be interested in fields generated over $\fd{Q}$ by algebraic numbers. 
Recall that an algebraic number is a root of an element of $\fd{Q}[x]$, the set of polynomials with rational coefficients. 
Given an algebraic number $\alpha$, the field it generates over $\fd{Q}$, denoted $\fd{Q}(\alpha)$, is defined to be the smallest set of complex numbers which includes both $\fd{Q}$ and $\alpha$, and which is closed under addition, subtraction, multiplication, and division. 
To describe its structure we need the minimal polynomial of $\alpha$. 
This is defined to be the lowest degree element of $\fd{Q}[x]$ (easily seen to be unique) having $\alpha$ as a root and leading coefficient $1$. 
Let \ea{
 P(x) = x^n + q_{n-1} x^{n-1} + \dots + q_0
 }
 be this polynomial, and consider the set
 \ea{
A &= \{ c_0 + c_1 \alpha + \dots c_{n-1} \alpha^{n-1} \colon c_0, c_1, \dots , c_{n-1} \in \fd{Q}\}.
 }
It is immediate that $\fd{Q} \subseteq A\subseteq \fd{Q}(\alpha)$, and that $A$ is closed under addition, subtraction. 
The fact that $\alpha^n = - q_0 - \dots - q_{n-1} \alpha^{n-1}$ means $A$ is closed under multiplication. 
With more effort it is possible to show that it is also closed under division. 
Consequently, $A = \fd{Q}(\alpha)$. 
Note that for given $u = c_0 + \dots c_{n-1} \alpha^{n-1}\in A$, the coefficients $c_j$ are unique. 
It follows that, just as one can identify the field $\fd{C}=\fd{R}(i)$ with the set of column vectors $\left(\begin{smallmatrix} c_0 \\ c_1\end{smallmatrix}\right) \in \fd{R}^2$, so one can identify the field $\fd{Q}(\alpha)$ with the set of column vectors $\left(\begin{smallmatrix} c_0 \\ \vdots \\ c_{n-1} \end{smallmatrix}\right) \in \fd{Q}^n$. 
The integer $n$ is called the degree of the field extension, denoted $[\fd{Q}(\alpha)\colon \fd{Q}]$. 

At this stage it will be helpful to consider some examples. 
Consider first the field $\fd{Q}(\alpha)$, where $\alpha = \mathrm{e}^{\frac{2\pi i}{m}}$ for some prime number $m>2$. 
The minimal polynomial of $\alpha$ is
$P(x) = x^{m-1} + x^{m-2} + \dots + 1$. It is easily seen that $f(\alpha^j) = 0$ for all $j$, meaning that $f$ completely factorizes over the field $\fd{Q}(\alpha)$:
\ea{
P(x) &= (x-\alpha) (x-\alpha^2) \dots (x-\alpha^{m-1})
}
A field $\fd{Q}(\alpha)$ with this property is said to be a normal, or Galois extension of $\fd{Q}$.

For an example of a field which is not a normal extension of $\fd{Q}$, consider $\alpha = 2^{\frac{1}{4}}$. 
The minimal polynomial is $P(x) = x^4-2$. 
It is easily seen to split into one quadratic and two linear factors over $\fd{Q}(\alpha)$:
\ea{
P(x) &= (x-\alpha) (x+\alpha) (x^2+\alpha^2)
}
It is impossible that $(x^2+\alpha^2)$ should factor over $\fd{Q}(\alpha)$ since its roots are $\pm i \alpha$, whereas $\fd{Q}(\alpha)$ is a subfield of the reals. 
It follows that $\fd{Q}(\alpha)$ is not a normal extension of $\fd{Q}$. 
It will be seen that the smallest field over which $P(x)$ completely factors is $\fd{Q}(i,\alpha)$. 
This too can be obtained by appending a single generator to $\fd{Q}$. 
Indeed, let $\beta = i + \alpha$. 
Clearly, $\fd{Q}(\beta) \subseteq \fd{Q}(i,\alpha)$. 
On the other hand, one finds, with the help of a package such as \emph{Magma} or \emph{Sage},
\ea{
i &= -\frac{1}{24}(5\beta^7 + 19\beta^5 + 5\beta^3 + 127\beta),
\nn
\alpha&= \beta - i = \frac{1}{24}(5\beta^7 + 19\beta^5 + 5\beta^3 + 151\beta).
}
So $\fd{Q}(\beta) = \fd{Q}(i,\alpha)$. 
With the the further assistance of a computer algebra package one finds that the minimal polynomial of $\beta$ is
\ea{
Q(x) &= x^8 + 4x^6 + 2x^4 + 28x^2 + 1.
}
It completely splits over $\fd{Q}(\alpha,i)=\fd{Q}(\beta)$:
\ea{
Q(x) &= (x-\alpha-i)(x+\alpha-i)(x-\alpha+i)(x-\alpha-i)
\nn
& \hspace{1.5 in} \times (x-i\alpha-i)(x+i\alpha-i)(x-i\alpha+i)(x-i\alpha-i).
}
We conclude that $\fd{Q}(\beta)$ is a normal extension of $\fd{Q}$. 
This calculation illustrates two important principles:
\begin{enumerate}
\item Given a set of $k$ generators $\alpha_1$, \dots, $\alpha_k$ it is always possible to find a single generator $\alpha$ such that $\fd{Q}(\alpha_1, \dots, \alpha_k)=\fd{Q}(\alpha)$.
\item Given an extension $\fd{Q}(\alpha)$ which is not normal, there always exists $\beta$ such that $\fd{Q}(\beta)$is normal and $\fd{Q}(\alpha) \subset \fd{Q}(\beta)$. The normal extension of smallest degree is called the normal closure of $\fd{Q}(\alpha)$.
\end{enumerate}

We next turn to the calculation of the Galois group of a normal field extension. 
Let $\fd{E} = \fd{Q}(\alpha)$ be normal over $\fd{Q}$, let $P(x) = x^n + c_{n-1} x^{n-1} + \dots + c_0$ be the minimal polynomial of $\alpha$, and let $\alpha_1$, \dots , $\alpha_n$ be is roots, with the indices chosen so that $\alpha_1 = \alpha$. 
Let $g$ be an element of $\gal(\fd{E}/\fd{Q})$. 
As with our discussion of $\gal(\fd{C}/\fd{R})$, it is completely determined by its action on the field generator $\alpha$. 
Acting with $g$ on both sides of
\ea{
\alpha^n + c_{n-1} \alpha^{n-1} + \dots + c_0 &= 0
}
we deduce 
\ea{
g(\alpha)^n + c_{n-1} g(\alpha)^{n-1} + \dots + c_0 &= 0.
}
implying $g(\alpha) = \alpha_j$ for some $j$. 
Conversely, for each $j$ there exists an automorphism $g\in \gal(\fd{E}/\fd{Q})$ such that $g(\alpha) = \alpha_j$. 
In short, the elements of $\gal(\fd{E}/\fd{Q})$ are in bijective correspondence with the roots of the minimal polynomial of the field generator. 
In particular, the order of $\gal(\fd{E}/\fd{Q})$ is the same as the degree $[\fd{E}\colon \fd{Q}]$.

We are now ready to describe the Galois correspondence between subfields and subgroups. 
Let $\fd{E} = \fd{Q}(\alpha)$ be a normal extension of $\fd{Q}$, and let $G=\gal(\fd{E}/\fd{Q})$ be the corresponding Galois group. 
Let $\mcl{L}_{\fd{E}}$ be the set of subfields $\fd{Q} \subseteq \fd{F} \subseteq \fd{E}$, and let $\mcl{L}_{G}$ be the set of subgroups $\la e \ra \subseteq H \subseteq G$. 
Let $\xi \colon \mcl{L}_{\fd{E}} \to \mcl{L}_{G}$ and $\eta \colon \mcl{L}_{G} \to \mcl{L}_{\fd{E}}$ be the maps defined by
\ea{
\xi(\fd{F}) &= \{ g \in G \colon g(u) = u \quad \forall u \in \fd{F}\},
\\
\eta(H) &= \{u \in \fd{E} \colon g(u) = u \quad \forall g \in H\}.
\label{eq:fixedField}
}
The field $\eta(H)$ is called the fixed field of $H$. 
It can be shown
\begin{enumerate}
\item $\xi$, $\eta$ are mutually inverse bijections.
\item $H$ is a normal subgroup of $G$ if and only if $\eta(H)$ is a normal extension of $\fd{Q}$.
\end{enumerate}

Finally, we have the famous result, that the elements of a field are all expressible in radicals if and only if the corresponding Galois group is solvable. 
Here, the statement that a number is expressible in radicals means that it can be built up from the rationals by means the four operations of addition, subtraction, multiplication and division together with the operation of taking roots. 
The statement that a group $G$ is solvable means that there exists a chain of subgroups
\ea{
H_0 = \la e \ra \subseteq H_1 \subseteq \dots \subseteq H_m = G
\label{eq:subNorm}
}
such that $H_j$ is a normal subgroup of $H_{j+1}$ with $H_{j+1}/H_j$ Abelian for $j=0,1 , \dots m-1$.

\section{The Galois Group of a Generic SIC Fiducial}
\label{sec:GaloisSIC}
One finds in practice that the matrix elements of every known generic SIC fiducial projector are expressible in radicals. 
This immediately tells us that the associated Galois group must be solvable. It turns out that one can say much more than that, as we now explain. 
It should be stressed that although we can prove some of the statements in this section, many of them are a matter of empirical observation. 
In a way that is what makes them interesting: They may be regarded as clues, the study of which may take us closer to a solution of the SIC-existence problem. 
For more details regarding the material in this section see refs.~\cite{Appleby:2013,Appleby:2017}.

Let $\Pi$ be a generic SIC fiducial projector, and let $\fd{E} = \fd{Q}(\Pi, \tau)$ be the field generated over the rationals by the standard basis matrix elements of $\Pi$ together with the number $\tau$. 
It can be shown that if $\Pi$, $\Pi'$ are two fiducials on the same $\PEC(d)$ orbit, then $\fd{Q}(\Pi, \tau)=\fd{Q}(\Pi', \tau)$. 
The field $\fd{E}$ is thus a feature of the orbit, not of the individual projector. 
We refer to it as the SIC field.

\begin{fact}
In every known case $\fd{E}$ is normal over $\fd{Q}$.
\end{fact}

\noindent This is a significant fact: For, as we saw in Section~\ref{sec:Galois}, there is no guarantee that a randomly chosen field extension will be normal. 

\begin{fact}
In every known case $\fd{E}$ is an extension of $\fd{K}=\fd{Q}(\sqrt{D})$, where $D$ is the square-free part of $(d-3)(d+1)$.
\end{fact}

\begin{fact}
In every known case $\gal(\fd{E}/\fd{K})$ is Abelian.
\end{fact}

\noindent Facts 2 and 3 tell us that the chain in Eq.~\eqref{eq:subNorm} can be chosen to be of length $3$,
\ea{
\la e \ra \subseteq H_1 = \gal(\fd{E}/\fd{K}) \subseteq G
}
with $H_1$ an index 2 subgroup of $G$. 
Loosely speaking, one may say that $\gal(\fd{E}/\fd{Q})$ is as close to Abelian as it could be, without actually being Abelian.

\vspace{0.05 cm} 
For later reference it will be convenient to introduce some terminology at this point. 
A field of the type $\fd{Q}(\sqrt{k})$ (respectively, $\fd{Q}(i\sqrt{k})$), with $k$ a square free integer greater than 1, is called a real quadratic field (respectively, imaginary quadratic field). 
If $\fd{E} \supset \fd{F}$ is such that $\gal(\fd{E}/\fd{F})$ is Abelian then we say that $\fd{E}$ is an Abelian extension of $\fd{F}$. 
Expressed in this language, the SIC field is an Abelian extension of a real quadratic field.

\begin{fact}
In every known case $\gal(\fd{E}/\fd{K})$ is the centralizer of complex conjugation.
\end{fact}

\noindent This tells us that if $g\in \gal(\fd{E}/\fd{K})$ then $g(\Pi)$ is another SIC fiducial (where by $g(\Pi)$ we mean the matrix obtained by acting with $g$ on the standard basis matrix elements of $\Pi$). 
Note that if $g$ does not commute with complex conjugation then $g(\Pi)$ is typically not even Hermitian, let alone a SIC fiducial projector. 

It is possible to make some strong statements about the structure of the group $\gal(\fd{E}/\fd{K})$. 
We first need to define two additional subfields. 
It often happens that different $\PEC(d)$ orbits generate the same SIC field. 
We refer to such groups of orbits as Galois multiplets. 
For example $4a$ is a Galois singlet, $9ab$ are a Galois doublet, $30abc$ are a Galois triplet and $21abcd$ are a Galois quartet (where the notation $9ab$ is shorthand for $9a$, $9b$, and where we employ the Scott-Grassl~\cite{Scott:2010a} labelling convention).
Given a fiducial projector $\Pi$ define $G_0$ to be the group consisting of all $g\in \gal(\fd{E}/\fd{K})$ such that $g(\Pi)$ is on the same $\PEC(d)$ orbit as $\Pi$. 
One finds in practice that $G_0$ only depends on the multiplet, and not on the individual projector. 
We refer to it as the group of orbit-fixing automorphisms. 
Let $\fd{E}_0$ be the fixed field of $G_0$. 

\begin{fact}
In every known case the index $[\fd{E}_0\colon \fd{K}]$ is the same as the size of the multiplet (i.e.\ 1 for a singlet, 2 for a doublet, etc).
\end{fact}

Let $g_c$ be complex conjugation, let $g$ be any element of $\gal(\fd{E}/\fd{Q})$ with the property $g(\sqrt{D}) = -\sqrt{D}$, and let $\bar{g}_1 = g g_c g^{-1}$. 
It is easily seen that $\bar{g}_1$ is (a) independent of the choice of $g$ and (b) order $2$. 
Let $\fd{E}_1$ be the fixed field of $\la \bar{g}_1 \ra$. 

\begin{fact}
In every known case $\fd{E}=\fd{E}_1(i\sqrt{d'})$.
\end{fact}

We thus have a tower 
\ea{
\fd{Q} \subseteq \fd{K} \subseteq \fd{E}_0 \subseteq \fd{E}_1 \subseteq \fd{E}.
}
Of the four groups $\gal(\fd{K}/\fd{Q}$, $\gal(\fd{E}_0/\fd{K})$, $\gal(\fd{E}_1/\fd{E}_0)$, $\gal(\fd{E}/\fd{E}_1)$, the group $\gal(\fd{E}_1/\fd{E}_0)$ is much the largest (except when $d$ is small). 
We now address the problem of characterizing its structure.

\begin{fact}
On every known $\PEC(d)$ orbit there exists a fiducial projector $\Pi$ such that
\ea{
\Tr(D_{\mbf{p}}\Pi) \in \fd{E}_1
\label{eq:olpFld}
}
for all $\mbf{p}$.
\end{fact}

\noindent A centred fiducial having this property is called strongly centred. 
If $d\neq 0 \mod 3$ every centred fiducial is strongly centred. 
If $d=0 \mod 3$ that is not the case, but the class of strongly centred fiducials is still non-empty. 
From now on it will be assumed without comment that the fiducial is strongly centred.

\begin{fact}
In every known case $\gal(\fd{E}_1/\fd{E}_0)$ permutes the overlaps. 
Specifically, there corresponds to each $g\in \gal(\fd{E}_1/\fd{E}_0)$ a matrix $G_g \in \GL(2,\fd{Z}/d'\fd{Z})$ such that
\ea{
g\big( \Tr(\Pi D_{\mbf{p}})\big) &= \Tr(\Pi D_{G_g \mbf{p}}) 
}
for all $\mbf{p}$.
\end{fact}

\noindent The matrix $G_g$ is not unique. 
Let 
 \ea{
 \stb(\Pi) &= \{ (\det F) F \colon F \in \stbz(\Pi)\}.
 }
Then $\stb(\Pi)$ is the overlap stability group: if $G\in \stb(\Pi)$ then $\Tr(\Pi D_{G\mbf{p}}) = \Tr(\Pi D_{\mbf{p}})$ for all $\mbf{p}$. 
One finds in practice that $G_g$ is always in the centralizer of $S(\Pi)$. 
Consequently we can replace $G_g$ with an arbitrary element of the coset $G_g S(\Pi)$. 

\begin{fact}
In every known case the map $g\to G_g$ defines an isomorphism
\ea{
\gal(\fd{E}_1/\fd{E}_0) &\cong M(\Pi) / \stb(\Pi)
}
where $M(\Pi)$ is a maximal Abelian subgroup of $\GL(2,\fd{Z}/d'\fd{Z})$ containing $\stb(\Pi)$. 
For a type-$z$ orbit there is only one such subgroup, namely the centralizer of $\stb(\Pi)$. 
For a type-$a$ orbit there are several possibilities, categorized in ref.~\cite{Appleby:2017}.
\end{fact}

In the Introduction we remarked that the numbers in the SIC problem, superficially regarded, seem to be devoid of any obvious pattern. 
The number-theoretic features adduced in this section may, perhaps, go some way towards dispelling that impression. 
Particularly striking are the fact that $\fd{E}$ is an Abelian extension of the real quadratic field $\fd{Q}(\sqrt{D})$, and the isomorphism $\gal(\fd{E}_1/\fd{E}_0) \cong M(\Pi) / \stb(\Pi)$.

\section{Mathematical Background: Hilbert's \texorpdfstring{$12^{\rm{th}}$}{12th} Problem}
\label{sec:Hilbert}
The study of Abelian extensions of a given number field has played a central role in the development of algebraic number theory. 
Kronecker began this line of investigation in the $19^{\rm{th}}$ century, by asking what is the general form of an Abelian extension of the rationals. 
The answer is provided by the Kronecker-Weber theorem~\cite{Cohn:1978}, which states that if $\fd{E}$ is an algebraic number field, then $\gal(\fd{E}/\fd{Q})$ is Abelian if and only if $\fd{E}$ is a subfield of $\fd{Q}(\mathrm{e}^{\frac{2\pi i}{n}})$, for some integer $n$. 

A similar result can be proved for Abelian extensions of imaginary quadratic fields. 
Here too Kronecker was responsible for posing the problem, and for many of the insights which eventually led to a solution, although the proof was not completed until long after his death~\cite{Vladut:1991}. 
The result states~\cite{Cohn:1978,Vladut:1991,Cox:1989} that, if $\fd{E}$ is an algebraic extension of an imaginary quadratic field, then the extension is Abelian if and only if $\fd{E}$ is a subfield of a field generated by special values of certain modular and elliptic functions. 

We thus know of two cases where a type of Abelian extension is generated by special values of transcendental functions. 
In his $12^{\rm{th}}$ problem Hilbert asked if the phenomenon generalizes. 
The attempt to answer that question has been one of the main foci of research in algebraic number theory ever since. 
However, although it has been very fruitful, in the sense that it has stimulated many important developments, the original problem remains essentially unsolved.

A solution to Hilbert's $12^{\rm{th}}$ problem for Abelian extensions of the field $\fd{K}$ requires two things:
\begin{enumerate}
\item Identifying a set of fields which play the role that the fields $\fd{Q}(\mathrm{e}^{\frac{2\pi i}{n}})$ do for $\fd{Q}$.
\item Identifying a function or functions special values of which generate the fields, analogous to the way in which a special values of the exponential function generate the fields $\fd{Q}(\mathrm{e}^{\frac{2\pi i}{n}})$.
\end{enumerate}
The first of these tasks has been accomplished. 
The fields in question are the ray class fields over $\fd{K}$, the number playing the role of $n$ being called the conductor\cite{Cohn:1978}. 
Moreover there are algorithms for calculating the ray class fields. 
What remains is the problem of finding suitable transcendental functions to generate the fields.

The relevance of all this to the SIC problem is that, after the rationals, and imaginary quadratic fields, the obvious next case to try is Abelian extensions of real quadratic fields. 
This is, of course, the type of field which features in the SIC problem. 
Moreover, it turns out that certain SICs generate the ray class fields which are relevant to Hilbert's $12^{\rm{th}}$ problem.

\section{SICs and Ray Class Fields}
\label{sec:SICsRayClass}
As we explain in this section, for each square-free positive integer $D$, infinitely many of the ray-class fields over $\fd{Q}(\sqrt{D})$ are generated by SICs. 

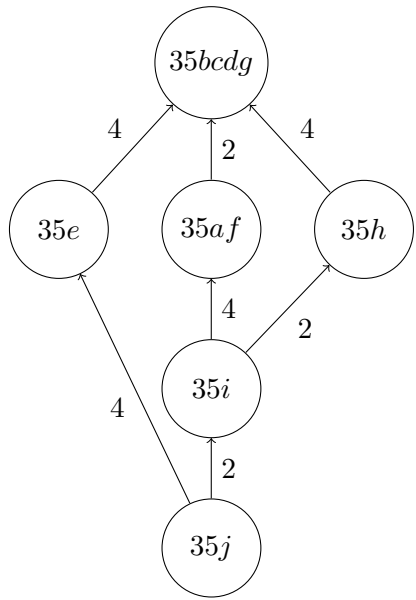
\begin{figure}[t!]
\begin{tikzpicture}
[auto, nd/.style={circle,minimum size = 13 mm,draw}]

\matrix[row sep=8mm,column sep=6mm] {
 &\node (35bcdg) [nd] {$35bcdg$};
\\
\node (35e) [nd] {$35e$}; & \node (35af) [nd] {$35af$}; & \node (35h) [nd] {$35h$};
\\
 &\node (35i) [nd] {$35i$};
\\
 &\node (35j) [nd] {$35j$};
\\
};
\draw[->] (35j) to node[swap] {2} (35i);
\draw[->] (35i) to node[swap] {2} (35h);
\draw[->] (35j) to node{4} (35e);
\draw[->] (35i) to node[swap] {4} (35af);
\draw[->] (35e) to node{4} (35bcdg);
\draw[->] (35af) to node[swap] {2} (35bcdg);
\draw[->] (35h) to node[swap] {4} (35bcdg);
\end{tikzpicture}
\caption{\label{fg:multiplet}Field inclusions for the six known SIC multiplets in dimension 35. 
The arrows run from the smaller field to the larger; numbers beside the arrows are the degrees of the extensions. 
The minimal multiplet is $35j$; the maximal multiplet is $35bcdg$.}
\end{figure}

In some dimensions there is only one known SIC multiplet. 
However, in many cases there are more~\cite{Scott:2010a,Appleby:2017}. 
When that happens one finds, in every known case,
\begin{enumerate}
\item Each multiplet generates a unique field, distinct from the fields generated by all the other multiplets,
\item There is a unique minimal multiplet, whose field is contained in the fields generated by all the other multiplets,
\item There is a unique maximal multiplet, whose field contains the the fields generated by all the other multiplets.
\end{enumerate}
The situation is illustrated in Fig.~\ref{fg:multiplet}, which shows the pattern of field inclusions for dimension $35$. 
It turns out that in every known case the minimal multiplet generates a ray-class field\footnote{There are actually four ray-class fields over $\fd{Q}(\sqrt{D})$ whose conductors have finite part $d'$. 
The SIC field $\fd{E}$ generated by a minimal multiplet is the largest of these fields; specifically, the one with ramification allowed at both infinite places. 
The other three are subfields of $\fd{E}$, and they also play a role in the theory. 
In particular, the field $\fd{E}_1$ defined earlier is a ray-class field. 
}
over $\fd{Q}(\sqrt{D})$ whose conductor has finite part $d'$. 

We have only confirmed this statement for 24 cases (specifically, the minimal multiplets in dimensions $4$--$21$ inclusive, $24$, $28$, $30$, $35$, $39$, and $48$). 
Suppose, however, it is true in every dimension greater than 3. Then it can be shown that
infinitely many of the ray-class fields over any given real quadratic field are generated in this way, by minimal SIC multiplets. 
Indeed, let $D$ be an arbitrary square-free positive integer. 
$D$ is the square-free part of $(d-3)(d+1)$ for some dimension $d$ if and only if
\ea{
(d-1)^2 - m^2 D = 4
}
for some positive integer $m$. 
This is a modified version of Pell's equation, and like Pell's equation it has infinitely many solutions~\cite{Cohn:1978,Appleby:2016}. 
If the solutions are written as an increasing sequence $d_1 < d_2 < \dots $, the first solution can be found by calculating a fundamental unit for the field $\fd{Q}(\sqrt{D})$, and the remaining solutions can be calculated in terms of that using 
\ea{
d_j &= 1+2 T_j\left(\frac{d_1-1}{2}\right)
}
where $T_j$ is a Chebyshev polynomial of the first kind. 
In this way we find, for example, that the sequence of dimensions for which $D=5$ is
\ea{
d=4, 8, 19, 48, 124, 323, 844, 2208, 5779, 15128, 39604, 103683, \dots
}
It will be observed that this contains subsequences such as 4, 8, 48, 2208, \dots or 4, 124, 15128, \dots or 19, 323, 103683, \dots in which each term is a divisor of the one following. 
We call such subsequences dimension towers. 
Their significance is that if $\fd{E}_1$, $\fd{E}_2$ are ray class fields over the same field $\fd{K}$, and if the conductor of $\fd{E}_1$ divides the conductor of $\fd{E}_2$, then $\fd{E}_1$ is a subfield of $\fd{E}_2$. 
One finds that this embedding of the fields in a dimension tower reflects some intriguing relationships between the underlying SICs. 
We are actively exploring these relationships.

In this section we have focussed on the minimal multiplet in each dimension. 
It is natural to ask if the other multiplets also have interesting number-theoretic properties. 
We have yet to investigate that question. 
The reason is that until the work reported in ref.~\cite{Appleby:2017} we only had four examples of two multiplets in a single dimension, and none of three or more. 
This was not a sufficiently large dataset for a computational investigation of the kind on which our results are mostly (though not entirely) based.

\section{Mathematical Background: Algebraic Integers and the Unit Group}
\label{sec:AlgInt}
The rational numbers are the set of all integer ratios:
\ea{
\fd{Q} &= \left\{ \frac{n}{m} \colon n,m\in \fd{Z}, m\neq 0\right\}.
}
It turns out that it is possible to carry through an analogous construction for an arbitrary algebraic number field. 
In fact, let $\fd{F}$ be such a field, and let $\fd{Z}_{\fd{F}}$ be the subset consisting of all elements of $\fd{F}$ whose minimal polynomials have integer coefficients. 
It is easily seen that $\fd{Z}_{\fd{F}} \cap \fd{Q} = \fd{Z}$. 
Indeed, if $u\in \fd{Q}$ then its minimal polynomial is $x-u$, which has integer coefficients if and only if $u \in \fd{Z}$. 
In particular, $\fd{Z}_{\fd{F}}$ contains $0$ and $1$. 
With a little more work it can be shown~\cite{Cohn:1978} that $\fd{Z}_{\fd{F}}$ is closed under addition and multiplication, so that $u+v$, $uv\in \fd{Z}_{\fd{F}}$ for all $u, v\in \fd{Z}_{\fd{F}}$. 
It is therefore a ring. 
One can also show
\ea{
\fd{F} &= \left\{ \frac{n}{m} \colon n,m\in \fd{Z}_{\fd{F}}, m\neq 0\right\}.
}
Thus, the ring $\fd{Z}_{\fd{F}}$ stands in the same relation to $\fd{F}$ that the ring $\fd{Z}$ does to $\fd{Q}$. 
Accordingly, its elements are called algebraic integers. 

In the same way that ordinary number theory studies the divisibility properties of ordinary integers, so algebraic number theory studies the divisibility properties of algebraic integers. 
In particular, it is possible to prove a prime factorization theorem for algebraic integers which includes the factorization theorem for ordinary integers as a special case. 

These considerations are intimately involved in the definition of the ray class fields~\cite{Cohn:1978} featuring in the last two sections. 
Let $\fd{E}$ be a ray class field over $\fd{K}$. 
It may happen that a prime in $\fd{Z}_{\fd{K}}$ ceases to be prime when lifted to $\fd{Z}_{\fd{E}}$. 
In that case one may ask how it factors in the larger ring. 
If it splits into a product of primes with multiplicity $1$ then it is said to be unramified in $\fd{Z}_{\fd{E}}$; otherwise it is said to ramify. 
The conductor of $\fd{E}$ provides information as to which primes ramify, and which do not. 

In the case of the ordinary integers every number has a unique prime factorization aside from $0$ and $\pm 1$. 
The significance of $0$ is that it is the additive identity. 
The significance of $\pm 1$ is that these are the non-zero integers $u$ such that $1/u$ is also an integer. 
The pair $\{1 , -1\}$ form a group under multiplication which is called the unit group. 
The unit group of an arbitrary algebraic number field $\fd{F}$ is defined in the same way; the only difference being that, whereas it is trivial to calculate the unit group for $\fd{Z}$, calculating it in the general case can be a difficult problem. 
The general structure is specified by Dirichlet's theorem~\cite{Cohn:1978}, which states that the group is isomorphic to
\ea{
(\fd{Z}/m\fd{Z}) \times \underbrace{ \fd{Z} \times \dots \times \fd{Z} }_{\text{$n$ copies of $\fd{Z}$}}.
\label{eq:UnitGp}
}
for suitable integers $m$, $n$. 
In other words, there exist $\sigma$, $u_1$, \dots, $u_n \in \fd{Z}_{\fd{F}}$ such that 
\begin{enumerate}
\item $\sigma$ is the $m^{\rm{th}}$ root of unity $\mathrm{e}^{\frac{2\pi i}{m}}$ (and therefore automatically a unit),
\item $u_1$, \dots, $u_n$ are units of infinite order,
\item An arbitrary unit $u$ can be written $u = \sigma^{r_0}_{\vphantom{1}} u_1^{r_1} \dots u_n^{r_n}$, for unique $r_0$, \dots, $r_n\in \fd{Z}$ with $r_0$ in the range $0 \le r_0 < m$. 
\end{enumerate}
Dirichlet's theorem also gives a formula for the integer $n$. 

There are algorithms for calculating the unit group. 
Using the implementation in \emph{Magma} one quickly ($t\ll 1 \ \mathrm{sec}$) finds, for example:
\begin{enumerate}
\item For the field $\fd{Q}(\sqrt{2})$ one has $m=2$, $n=1$, a possible choice for $u_1$ being
\ea{
u_1 &= \sqrt{2}+1,
}
\item For the field $\fd{Q}(\mathrm{e}^{\frac{2\pi i}{7}})$ one has $m=14$, $n=2$, possible choices for $u_1$, $u_2$ being
\ea{
u_1 &= 2\cos\frac{\pi}{7}, & u_2 &= 2\cos\frac{4\pi}{7}.
}
\end{enumerate}
However, for fields with larger discriminant the calculation is too slow to be practical (as it happens this is a problem for which a quantum computer would give an exponential speed-up~\cite{Hallgren:2005, Hallgren:2007, Eisentrager:2014}).

\section{SICs and the Unit Group}
\label{sec:SICsUnit}
It appears from the facts adduced in Section~\ref{sec:SICsRayClass} that SICs in some sense ``know'' about the way in which primes ramify in the field $\fd{E}$. 
It will appear from the facts adduced in this section that they also ``know'' about the unit group in the field $\fd{E}_1(\sqrt{d+1})$. 

Let $\Pi$ be a strongly centred, generic SIC fiducial in dimension $d$, and let $\mathrm{e}^{i\theta_{\rm{p}}}$ be the overlap phases, as defined by 
Eq.~\eqref{eq:olpDef}. 
It follows from Eq.~\eqref{eq:olpFld} that $\mathrm{e}^{i\theta_{\rm{p}}} \fd{E}_1 (\sqrt{d+1})=\fd{F}$. 
In many cases $\sqrt{d
+1} \in \fd{E}_1$, so that $\fd{F}= \fd{E}_1$; however, that is not always the case. 
One finds, for every known case, that the $\mathrm{e}^{i
\theta_{\mathrm{p}}}$ are algebraic integers. 
Since the multiplicative inverse of a phase is the same as its complex conjugate, and since 
the conjugate of an algebraic integer is another algebraic integer, it follows that the $\mathrm{e}^{i\theta_{\rm{p}}}$ are units. 
Define $U^{\rm{p}}$ to be the subgroup of the unit group consisting of those units which are also phases, and $U^{\rm{o}}$ to be the subgroup of $U^{\rm{p}}$ which is generated by the overlap phases together with $-1$ (for an illustration in the case of orbit $4a$, see Bengtsson's contribution to this volume). 
It is, of course, guaranteed that $U^{\rm{o}}$ is a subgroup of $U^{\rm{p}}$. 
What is much less obvious is the specific way it embeds. 
In ref.~\cite{Appleby:2016} we calculated the unit group for the seven orbits $4a$, $5a$, $6a$, $7b$, $8b$, $12b$ and $19e$. 
In every case we found 
\begin{equation}
\text{either} \qquad U^{\rm{p}} = U^{\rm{o}} \qquad \text{or} \qquad U^{\rm{p}} = U^{\rm{o}} \times U^{\rm{c}},
\label{eq:unitAB}
\end{equation}
where $U^{\rm{c}}$ is another subgroup having the same rank as $U^{\rm{o}}$. 
Unfortunately, the difficulties mentioned at the end of the last section, meant that it was impractical to calculate $U^{\rm{p}}$ in any more cases. 
However, we did calculate $U^{\rm{o}}$ (which only depends on a knowledge of the overlaps) in a further ten cases. 
We also derived a formula for the rank of $U^{\rm{p}}$. 
In this way we were able to confirm that in these additional cases, $\Rank (U^{\rm{o}})$ is equal either to $\Rank(U^{\rm{o}})$ or to $\frac{1}{2}\Rank(U^{\rm{o}})$. 
This is consistent with the conjecture that Eqs.~\eqref{eq:unitAB} are generally valid.
 
The fact that the overlap phases are units, and that they generate a subgroup of the unit group with such strikingly simple properties, is yet another illustration of the point, that the numbers featuring in the SIC problem appear to have some very special properties when regarded from a number-theoretic point of view.

\section{Conclusion}
\label{sec:conc}
If SICs existed in every finite dimension, that would be saying something important about the geometry of quantum state space. Similarly, a solution to a Hilbert's twelfth problem for real quadratic fields would be deeply important for number theory. 
It is remarkable, and to our mind very satisfying, that these two problems, ostensibly so different from one another, should appear on closer examination to be intimately connected. 

Our original reason for embarking on this research was the hope that it would lead to a solution to the SIC existence problem. 
That remains an important motivation. 
For instance, in Section~\ref{sec:SICsRayClass} we described dimension towers: i.e.\ sequences such as $4, 8, 48, \dots$ giving rise to the same integer $D$, and such that each successive dimension is a multiple of the one before. 
As we noted this means the corresponding fields embed. 
Moreover, there are some intriguing relationships between the SICs themselves in these dimensions. 
This raises the question, whether it might be possible to give an inductive proof, by showing that SIC existence at one level of a tower implies existence at the next. 
Of course, the idea of trying to prove existence inductively is obvious, and has probably occurred to everyone who has thought seriously about the problem. 
However, it is hard to see how anyone could have been led to consider sequences like $4, 8, 48, \dots$, or $4, 124, 15128, \dots$ if it were not for the clue from number theory.

The facts adduced in this paper, assuming they generalize, mean that a constructive proof of SIC existence needs to accomplish much more than was previously realized. 
Not only does it need to explain a geometric feature of quantum state space. 
It also needs to explain
\begin{enumerate}
\item Why SICs generate number fields with such very special properties,
\item Why the overlap phases are always algebraic integers,
\item Why the overlap phases, together with $-1$, always generate a subgroup of the unit group which is either identical to $U^{\rm{o}}$, or else is a direct summand of $U^{\rm{o}}$ having half the rank.
\end{enumerate}
This is a challenge. 
It is also, perhaps, a significant clue. 

So far we have been discussing the relevance of algebraic number theory to the SIC problem. 
However, it is possible that the connection will work both ways. 
For instance, long before we became interested in the number theoretic features of a SIC, it occurred to us, as it has doubtless occurred to many others, that proving SIC existence might reduce to proving a set of special function identities. 
Suppose that were the case. 
Then it might tell us which particular transcendental functions are needed for a solution to Hilbert's twelfth problem in the case of real quadratic fields.

\section*{Acknowledgements}
We are grateful to John Coates, Brian Conrad, Steve Donnelly, James McKee, Andrew Scott, and Chris Smyth for many useful comments and discussions. 
This research was supported in part by the Australian Research Council via EQuS project number CE11001013, and in part by Perimeter Institute for Theoretical Physics. 
Research at Perimeter Institute is supported by the Government of Canada through Innovation, Science and Economic Development Canada and by the Province of Ontario through the Ministry of Research, Innovation and Science. 
SF acknowledges support from an Australian Research Council Future Fellowship FT130101744 and JY acknowledges support from National Science Foundation Grant No. 116143.

\end{document}